%% file: prab_cupid.tex
\documentclass[%
 amsmath,amssymb,
 aps,
prab,
twocolumn,
]{revtex4-2}

\usepackage{graphicx}
\usepackage{dcolumn}
\usepackage{bm}

\usepackage{siunitx}
\usepackage{comment}
\usepackage{layouts}
\usepackage{upgreek}
\usepackage{tikz}
\usepackage[USenglish]{babel}
\usepackage{makecell}
\usepackage{url}
\usepackage{hyperref}

\begin{document}


\title{\textbf{Nanosecond Radio-Frequency Pulse Driven Photogun for Very Hard X-ray Free-electron Laser} 
}%

\author{Wei Hou Tan}\email{whtan@slac.stanford.edu}
\author{River R. Robles}
\author{Juan E. Hernandez Jr.}
\author{Emilio Alessandro Nanni}
\author{Ankur Dhar}\email{adhar@slac.stanford.edu}
\affiliation{SLAC National Accelerator Laboratory, 2575 Sand Hill Rd, Menlo Park, California 94025, USA}
\date{\today}

\begin{abstract}
One pathway to producing high brightness electron beams is to use a radio-frequency (rf) driven high field photogun to rapidly accelerate photoemitted electrons to the relativistic regime and preserve the brightness. However, the highest attainable field is limited by rf breakdowns of materials used in a photogun. Shortening rf pulse duration feeding into a photogun provides a viable pathway to achieve high field and prevent rf breakdowns.
Here we propose and investigate the Compressed Ultrashort Pulse Injector Demonstrator (CUPID), a nanosecond rf pulse driven photogun powered by a klystron and rf pulse compression system capable of achieving \SI{300}{\mega\watt} at \SI{20}{\nano\second} duration, to produce bright electron beams with high electric field. We first introduce the design of the CUPID photogun and its expected rf performance at \SI{500}{\mega\volt/\metre} driven by high power nanosecond rf pulses, followed by beam dynamics studies showing its capability for producing bright electron beams with \SI{60}{\nano\metre} emittance when forming a photoinjector with a superconducting solenoid and downstream accelerating structures. Finally, we show a proof-of-concept start-to-end simulation of the CUPID photoinjector paired with the existing Linac Coherent Light Source (LCLS) copper accelerator free-electron laser (FEL) to demonstrate achievable \si{\milli\joule} pulse energy very hard x-ray photons at \SI{40}{\kilo\electronvolt} or higher.
\end{abstract}

\maketitle

\section{\label{sec:intro}Introduction}
High brightness photoguns have a wide range of scientific applications based on electron accelerators, such as free-electron lasers (FEL) \cite{osti_1616511}. The launch of the world's first hard x-ray FEL at the Linac Coherent Light Source (LCLS) of SLAC National Accelerator Laboratory opened a new frontier of photon science and found numerous applications in a wide range of research disciplines, such as biology, chemistry and material sciences \cite{emma_first_2010,RevModPhys.88.015007}. These scientific breakthroughs in turn drive the development of x-ray FELs to push for even higher photon and pulse energies, such as x-ray photons at \SI{40}{\kilo\electronvolt} or higher, to enable newer scientific studie[s \cite{osti_1183982,PhysRevB.89.184105,Sheffield:FEL2017-MOD06,instruments3040052,osti_1835032,euxfel_hardxray}.

\begin{figure}[h!]
\includegraphics{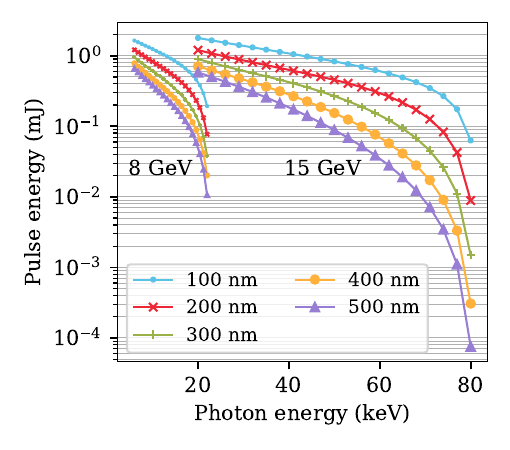}
\caption{\label{fig:fel_pulse_energy} X-ray FEL outputs for different photon energies with \SI{8}{\giga\electronvolt} and \SI{15}{\giga\electronvolt} \SI{100}{\pico\coulomb} \SI{3.5}{\kilo\ampere} beams with emittances at \num{100}, \num{200}, \num{300}, \num{400} and \SI{500}{\nano\metre} respectively, calculated using Ming-Xie parameterization. Parameters taken from Fig.~1-3 of ref \cite{osti_1616511}.}
\end{figure}

The performance of a FEL is proportional to the brightness of electron beams, as shown in the relation
\begin{align}
\rho &\propto\mathcal{B}_\text{6D}=\frac{2I}{\varepsilon_n^2\sigma_\gamma}\;,
\end{align}
where $\rho$ is the Pierce parameter that characterizes the efficiency of a FEL and $\mathcal{B}_\text{6D}$ is the six-dimensional beam brightness, $I$ is the beam current, $\varepsilon_n$ is the beam normalized transverse emittance and $\sigma_\gamma$ is the normalized rms energy spread in terms of the Lorentz factor respectively \cite{osti_1616511,rosenzweig_ultra-compact_2020}.
In the past, more efforts were often devoted to minimizing the beam's transverse emittance than minimizing the energy spread. The rationale was that transverse emittance was much harder to achieve for FEL requirements, whereas the energy spread was smaller than necessary to suppress instability. For example, LCLS has a device called a laser heater to deliberately increase the beam's energy spread to suppress shot-noise-induced instability \cite{huang_suppression_2004}.
Hence, one often has to minimize the beam's transverse emittance to improve the beam's brightness. Lower emittance beams allow FELs to output higher x-ray photon energy and increase total pulse energy. Figure \ref{fig:fel_pulse_energy} shows total pulse energy versus x-ray photon energy for \SI{100}{\pico\coulomb} \SI{3.5}{\kilo\ampere} beam at \SI{8}{\giga\electronvolt} and \SI{15}{\giga\electronvolt} with transeverse emittances from \SI{100}{\nano\metre} to \SI{500}{\nano\metre}, calculated using Ming-Xie parameterization \cite{Xie:1995kp}. Beam parameters taken from Fig.~1-3 of ref \cite{osti_1616511}.

The beam brightness is directly influenced by the strength of the applied electric field in a photogun, where a radiofrequency (rf) driven photogun is usually employed to accelerate photoemitted electron beams to high energies to suppress the degradation caused by charge repulsion \cite{fraser-1986-a,CARLSTEN1989313}. The relation of beam brightness and the gradient of applied field is given by $\mathcal{B}_\text{6D}\propto E^\nu$, where $\nu$ is a parameter that depends on the beam shape. For example, for beams with small transverse-to-longitudinal aspect ratio, $\mathcal{B}_\text{6D}\propto E^2$ \cite{kim-1989-a,rosenzweig_charge_1995,bazarov-2009-a,filippetto-2014-a,rosenzweig_ultra-compact_2020}.

However, the use of high gradient electric fields causes rf breakdowns, where uncontrolled field emission of electrons from the inner surface of the gun occurs in large quantities and prevents the build-up of accelerating fields inside the gun \cite{doebert-2007-a,wang-2009-a,Dolgashev-2010-a,Laurent-2011-a}. Accordingly, the phenomenological model of the breakdown rate is proportional to the applied field gradient but inversely proportional to the rf pulse duration \cite{grudiev-2009-a}.
Increasing the operating frequency allows shorter rf pulse duration, as proposed in various photogun designs at higher frequencies \cite{limborg-2016-a,Graves:2017cqi,marsh-2018-a,rosenzweig_ultra-high_2018,gonzalez-iglesias_x-band_2021,giribono_dynamics_2023}.
This presents the possibility of achieving a high gradient photogun fed by ultrashort high power rf pulses. This approach was recently demonstrated at the Argonne Wakefield Accelerator (AWA), where an X-band \SI{11.7}{\giga\hertz} gun was operated at $\sim$\SI{400}{\mega\volt/\metre} without significant breakdowns \cite{tan_demonstration_2022}. AWA achieved this by using a wakefield accelerator that accelerates \SI{400}{\nano\coulomb} electron bunches to high energies and generates $\sim$\SI{9}{\nano\second} high power rf pulses from a power-extraction-and-transfer structure (PETS) \cite{shao-2019-a}. However, this approach requires a larger physical footprint and additional stability control to operate a dedicated wakefield accelerator.

Here we present Compressed Ultrashort Pulse Injector Demonstrator (CUPID), a \num{1.6} cell photogun that can achieve high gradient with ultrashort high power rf pulses generated by a novel ultra-high power rf pulse compressor instead of the wakefield accelerator approach of AWA. This photogun operates at \SI{11.424}{\giga\hertz} to utilize an available X-band klystron of the same frequency and rf pulse compressor recently commissioned at SLAC~\cite{dhar:ipac2025-wepb100}.

This paper reports our design work on CUPID for generating bright electron beams that can drive very hard x-rays at the LCLS FEL. We first introduce the design of the CUPID photogun capable of achieving \SI{500}{\mega\volt/\metre} on cathode with nanoseconds long rf pulses from an existing klystron and recently commissioned rf pulse compressor, operating at \SI{11.424}{\giga\hertz}. We present the rf and mechanical design of the CUPID photogun and its performance. After that, we show the beam dynamics studies of the CUPID photoinjector, where the photogun is paired with a dedicated solenoid, followed by downstream accelerating structures, to demonstrate its capability in generating bright electron beams, down to \SI{60}{\nano\metre} transverse emittance. We then demonstrate start-to-end simulations of the CUPID photoinjector with the existing LCLS copper accelerator to show achievable \si{\milli\joule} pulse energy up to \SI{60}{\kilo\electronvolt} x-ray.

\section{\label{sec:cupid_design}CUPID Photogun RF Design}
\subsection{Design}
The CUPID photogun is a 1.6 cell photogun that has a very fast filling time to attain a very high gradient electric field with a very short time duration to suppress rf breakdowns. This is achieved by designing the CUPID photogun to be over-coupled to have a very low loaded quality factor,
\begin{align}
    Q_l=\pi ft_\text{fill}\,,
\end{align}
where $Q_l$ is the loaded quality factor, $f$ is the frequency, $t_\text{fill}$ is the rf filling time.
For example, the AWA X-band photogun has $\sim\SI{5}{\nano\second}$ filling time from its high power short pulses \cite{chen2025shortpulsedrivenradiofrequencyxband}. In comparison, typical X-band photoguns have filling time in the order of hundreds of nanoseconds, such as the CompactLight project X-band photogun at $\sim$\SI{112}{\nano\second} \cite{gonzalez-iglesias_x-band_2021}.

Figure \ref{fig:cupid_cutview} shows a cut-away view of the CUPID photogun design.
The exit of the photogun is designed as a \SI{11.43}{\milli\metre} radius waveguide to connect to a TM$_{01}$ mode launcher.
A photo of a fabricated prototype cavity being used for upcoming high power tests is also shown in Fig.~\ref{fig:cupid_built_gun}.
Four tuning pins are available for frequency tuning of both cells.
The cathode for the CUPID photogun is a copper backplate.
It has large aperture sizes to reach a very low $Q_l$.
\begin{figure}
\tikzset{every picture/.style={line width=0.75pt}} 
\begin{tikzpicture}[x=0.75pt,y=0.75pt,yscale=-1,xscale=1]

\draw (247.25,140.28) node  {\includegraphics[width=174.37pt,height=195.43pt]{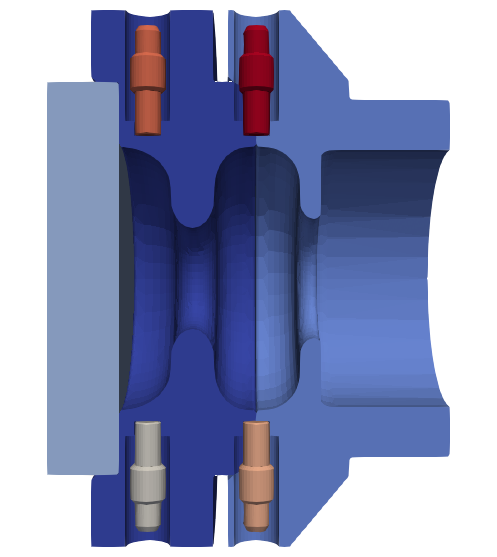}};
\draw [color={rgb, 255:red, 126; green, 211; blue, 33 }  ,draw opacity=1 ]   (339.35,85.67) -- (339.35,145.67) ;
\draw [shift={(340,147.67)}, rotate = 269.4] [color={rgb, 255:red, 126; green, 211; blue, 33 }  ,draw opacity=1 ][line width=0.75]    (10.93,-3.29) .. controls (6.95,-1.4) and (3.31,-0.3) .. (0,0) .. controls (3.31,0.3) and (6.95,1.4) .. (10.93,3.29)   ;
\draw [shift={(339.33,83.67)}, rotate = 89.4] [color={rgb, 255:red, 126; green, 211; blue, 33 }  ,draw opacity=1 ][line width=0.75]    (10.93,-4.9) .. controls (6.95,-2.3) and (3.31,-0.67) .. (0,0) .. controls (3.31,0.67) and (6.95,2.3) .. (10.93,4.9)   ;
\draw [color={rgb, 255:red, 126; green, 211; blue, 33 }  ,draw opacity=1 ]   (221.38,119.67) -- (221.38,144.33) ;
\draw [shift={(222,146.33)}, rotate = 268.67] [color={rgb, 255:red, 126; green, 211; blue, 33 }  ,draw opacity=1 ][line width=0.75]    (10.93,-3.29) .. controls (6.95,-1.4) and (3.31,-0.3) .. (0,0) .. controls (3.31,0.3) and (6.95,1.4) .. (10.93,3.29)   ;
\draw [shift={(221.33,117.67)}, rotate = 88.67] [color={rgb, 255:red, 126; green, 211; blue, 33 }  ,draw opacity=1 ][line width=0.75]    (10.93,-4.9) .. controls (6.95,-2.3) and (3.31,-0.67) .. (0,0) .. controls (3.31,0.67) and (6.95,2.3) .. (10.93,4.9)   ;
\draw [color={rgb, 255:red, 126; green, 211; blue, 33 }  ,draw opacity=1 ]   (187.33,147.67) -- (310.67,147.67) -- (338,147.67) ;
\draw [shift={(340,147.67)}, rotate = 180] [color={rgb, 255:red, 126; green, 211; blue, 33 }  ,draw opacity=1 ][line width=0.75]    (10.93,-3.29) .. controls (6.95,-1.4) and (3.31,-0.3) .. (0,0) .. controls (3.31,0.3) and (6.95,1.4) .. (10.93,3.29)   ;
\draw [shift={(185.33,147.67)}, rotate = 0] [color={rgb, 255:red, 126; green, 211; blue, 33 }  ,draw opacity=1 ][line width=0.75]    (10.93,-4.9) .. controls (6.95,-2.3) and (3.31,-0.67) .. (0,0) .. controls (3.31,0.67) and (6.95,2.3) .. (10.93,4.9)   ;
\draw [color={rgb, 255:red, 126; green, 211; blue, 33 }  ,draw opacity=1 ]   (278,113.67) -- (278,147.67) ;
\draw [shift={(278,147.67)}, rotate = 269.4] [color={rgb, 255:red, 126; green, 211; blue, 33 }  ,draw opacity=1 ][line width=0.75]    (10.93,-3.29) .. controls (6.95,-1.4) and (3.31,-0.3) .. (0,0) .. controls (3.31,0.3) and (6.95,1.4) .. (10.93,3.29)   ;
\draw [shift={(278,113.67)}, rotate = 89.4] [color={rgb, 255:red, 126; green, 211; blue, 33 }  ,draw opacity=1 ][line width=0.75]    (10.93,-4.9) .. controls (6.95,-2.3) and (3.31,-0.67) .. (0,0) .. controls (3.31,0.67) and (6.95,2.3) .. (10.93,4.9)   ;
\draw (342,107.33) node [anchor=north west][inner sep=0.75pt]   [align=left] {\SI{11.43}{\milli\metre}};
\draw (220.67,152) node [anchor=north west][inner sep=0.75pt]  [color={rgb, 255:red, 255; green, 255; blue, 255 }  ,opacity=1 ] [align=left] {\SI{29.26}{\milli\metre}};
\draw (229.33,124) node [anchor=north west][inner sep=0.75pt]  [color={rgb, 255:red, 255; green, 255; blue, 255 }  ,opacity=1 ] [align=left] {\SI{4.5}{\milli\metre}};
\draw (280.33,124) node [anchor=north west][inner sep=0.75pt]  [color={rgb, 255:red, 255; green, 255; blue, 255 }  ,opacity=1 ] [align=left] {\SI{5.3}{\milli\metre}};
\end{tikzpicture}
\caption{\label{fig:cupid_cutview}Cut-away view of a solid model of the CUPID photogun. Copper parts are shown in varying shades of blue. The exit of the photogun (right) is designed as a \SI{11.43}{\milli\metre} radius waveguide to connect to a TM$_{01}$ mode launcher.
Four tuning pins are available for frequency tuning of both cells.
The copper backplate (left) of the photogun serves as the cathode.
It has large aperture sizes to reach a very low $Q_l$.}
\end{figure}
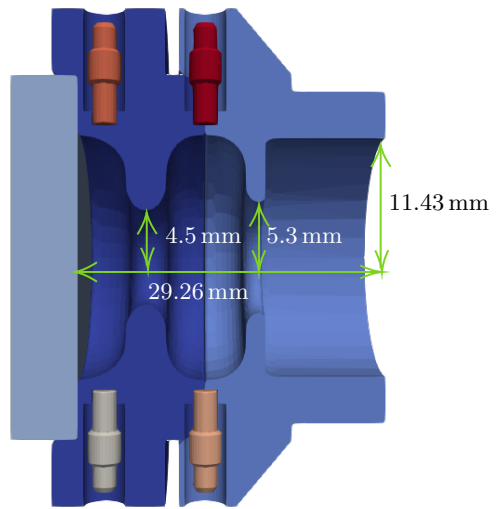
\begin{figure}[h!]
\includegraphics[width=.775\columnwidth]{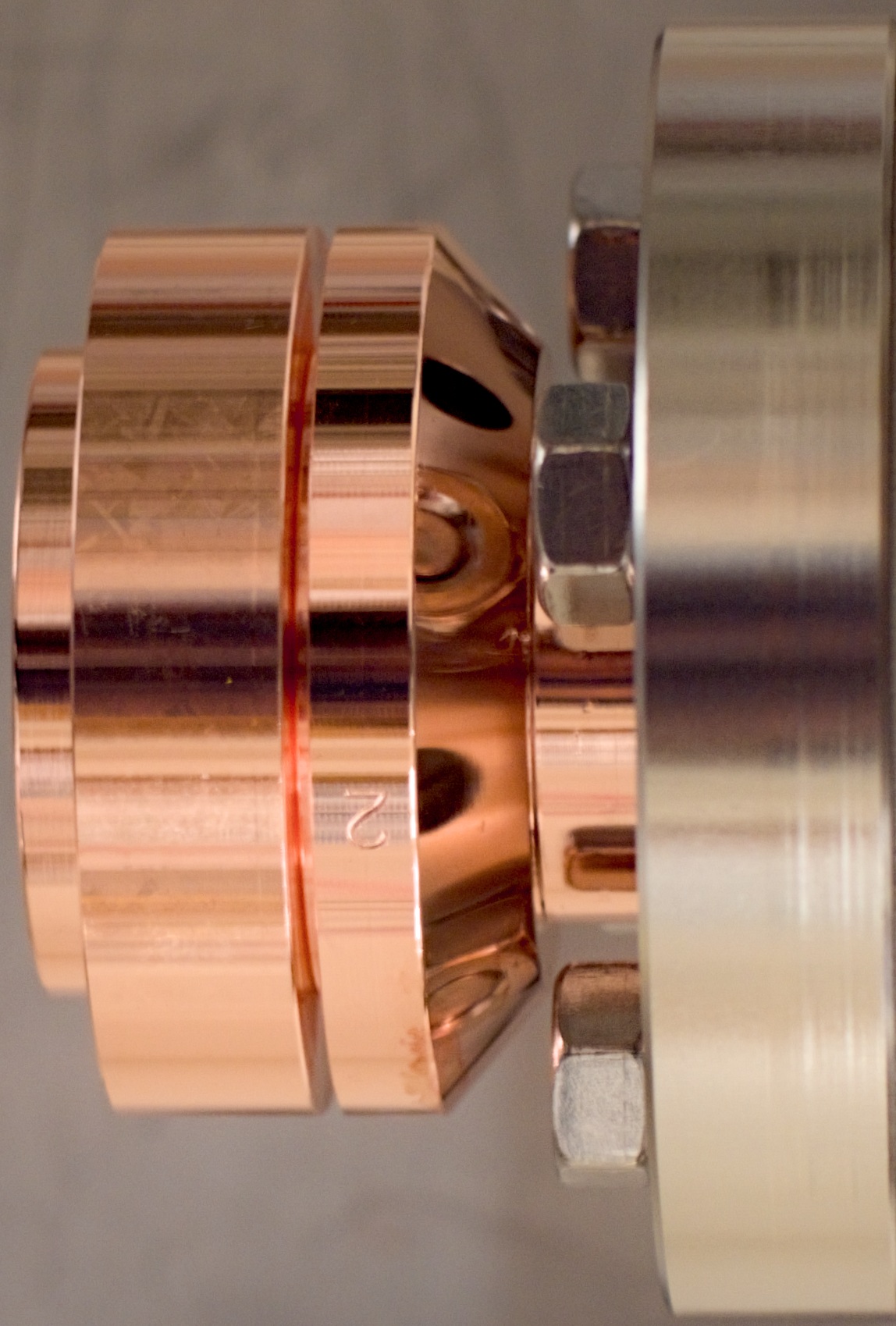}
\caption{\label{fig:cupid_built_gun} Photograph of fabricated test cavity for CUPID. This cavity will be used in upcoming high power tests to verify the ability to generate high electric fields with short pulses and low breakdown rate.}
\end{figure}

\subsection{Rf Performance}
\begin{figure}[h!]
\includegraphics[width=\linewidth]{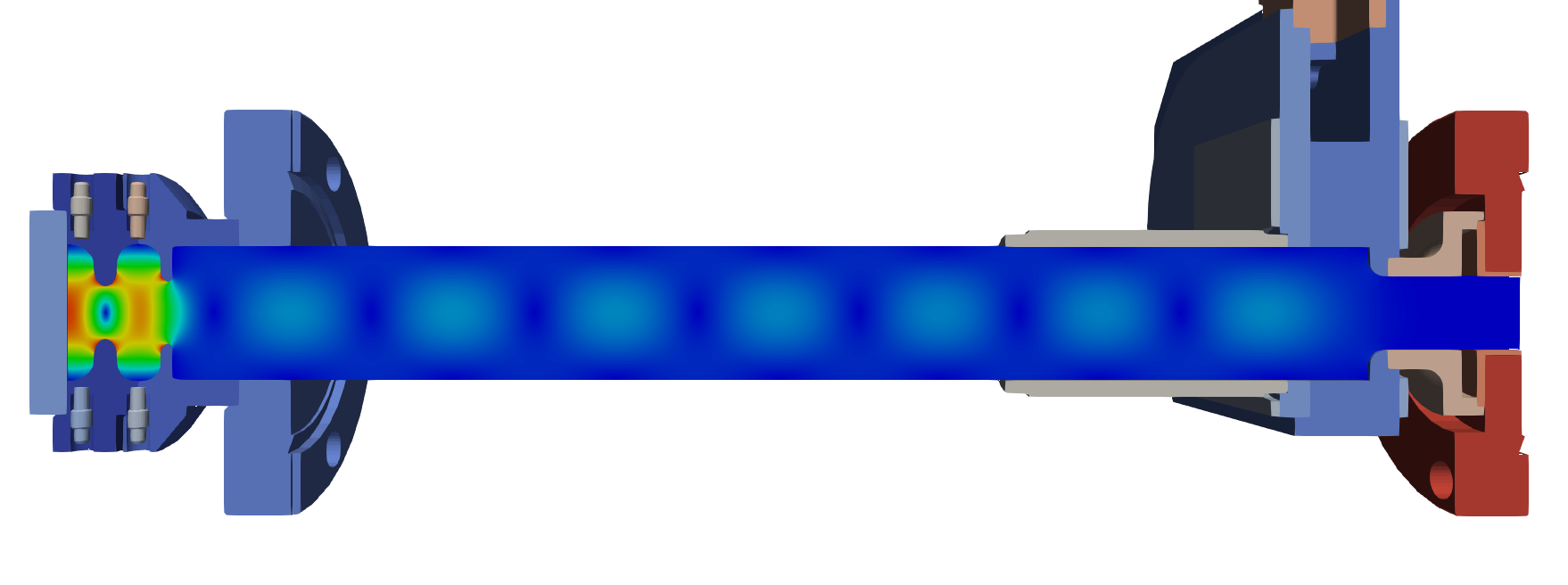}
\caption{\label{fig:cupid_mode_launcher}Cut-away view of the relative spacing between the CUPID photogun, the connecting waveguide, and the TM\textsubscript{10} mode launcher, with superimposed $\uppi$-mode field map. The relatively long waveguide connecting both components is a design choice to accommodate the physical size of a solenoid designed for CUPID photoinjector. Extended waveguide and solenoid are not shown in this view.}
\end{figure}

\begin{table}[b]
\caption{\label{tab:rf} Rf parameters for the CUPID photogun}
\begin{ruledtabular}
\begin{tabular}{lcc}
\textrm{Parameter}&
\textrm{Value}&
\textrm{Unit}\\
\colrule
                    Frequency& \num{11.424} & \si{\giga\hertz} \\
                    $Q_l$& \num{215} & - \\
                    $Q_0$& \num{6585} & - \\
                    Coupling coefficient $(\beta)$ & \num{29.63} & -\\
                    $S_{11}$ & \num{-0.59} & \si{\decibel}\\
                    Peak input power & \num{275} & \si{\mega\watt}\\
                    Field gradient at cathode & \num{500}& \si{\mega\volt/\metre}\\
\end{tabular}
\end{ruledtabular}
\end{table}
The electric field gradient at the cathode is designed to operate at \SI{500}{\mega\volt/\metre} fed by high power nanosecond rf pulses. This is achieved by using rf pulse compressors in conjunction with a high power klystron rf source. The \SI{500}{\mega\volt/\metre} gradient is chosen to match the normalized field strength parameter $\alpha_\mathrm{rf}=eE\lambda/4\pi mc^2\approx 2$ of the LCLS S-band \SI{120}{\mega\volt/\metre} photogun \cite{PhysRevSTAB.11.030703,kim-1989-a}.
For initial tests, the CUPID photogun is planned to connect to the available TM$_{01}$ X-band mode launcher \cite{osti_1336365,Torrisi_2019}.
Figure \ref{fig:cupid_mode_launcher} shows the cut-away view of the CUPID photogun, the connecting waveguide and the TM\textsubscript{01} mode launcher, with the $\uppi$-mode field map. The relatively long waveguide connecting both components is a design choice to accommodate the physical size of a solenoid designed for CUPID photoinjector.
Design studies of the CUPID photogun were conducted using the commercial software Ansys HFSS and cross-checked with multiphysics software suite ACE3P \cite{ansys,ace3p_2019}. Visualizations for ACE3P were performed using Paraview \cite{ahrens_paraview:_2005}.

\begin{figure}[h!]
\includegraphics[width=\linewidth]{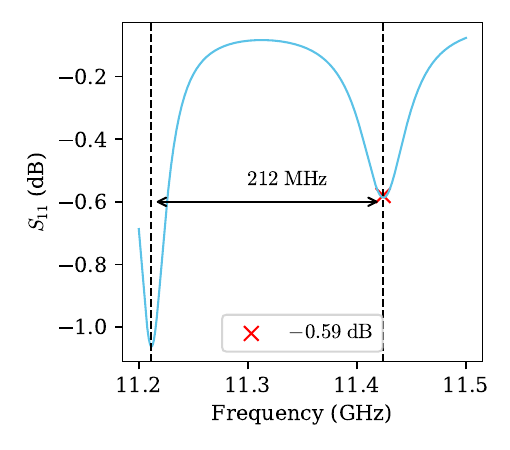}
\caption{\label{fig:sparam}Plot of simulated $S_{11}$ reflection parameter of the CUPID photogun. At \SI{11.424}{\giga\hertz}, the $S_\text{11}$ reflection parameter is \SI{-0.59}{\decibel} (``$\times$" sign).}
\end{figure}

Rf parameters are summarized at Tab.~\ref{tab:rf}. The frequency separation between $\uppi$-mode at \SI{11.424}{\giga\hertz} and the next nearest mode is \SI{212}{\mega\hertz}, which is larger than the \SI{100}{\mega\hertz} bandwidth of rf pulse compression network that is connected to the CUPID photogun. This means compressed pulses centered on \SI{11.424}{\giga\hertz} will not excite the O-mode. Figure \ref{fig:sparam} shows the plot of simulated $S_{11}$ reflection parameter of CUPID photogun. With a small $S_\text{11}$ reflection parameter at \SI{-0.59}{\decibel}, more than 95 \% of rf power will be reflected back from the CUPID photogun in the steady state. Hence, there are non-negligible fields present in the waveguide connecting the CUPID photogun and the TM\textsubscript{01} mode launcher as shown in Fig.~\ref{fig:cupid_mode_launcher}. This is a design choice to work with an existing mode launcher and alternatives will be investigated in the future. Fields present in the waveguide will have minimal impacts on the beam dynamics of emitted electrons, which will be discussed in Section \ref{sec:cupid_beam_dynamics}.

\subsection{Time Domain Simulations}
\begin{figure}[h!]
\includegraphics[width=\linewidth]{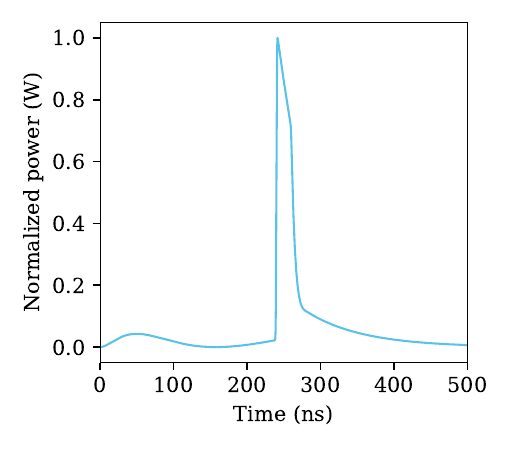}
\caption{\label{fig:pulse_time_domain}Plot of normalized power data from rf pulse compressors to be used for the CUPID photogun. The orange shaded region is the time period used for time domain simulations of CUPID photogun~\cite{dhar:ipac2025-wepb100}}
\end{figure}

\begin{figure}[h!]
\includegraphics[width=\linewidth]{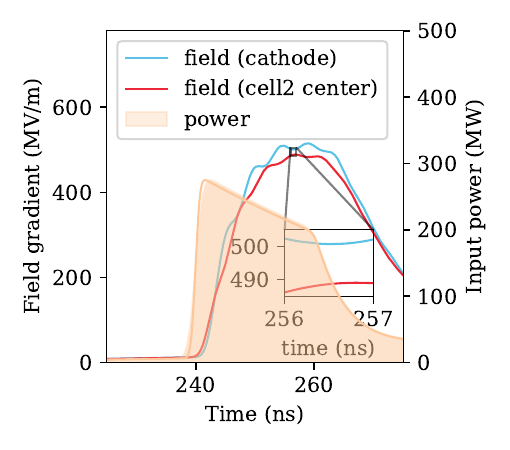}
\caption{\label{fig:field_pulse}Above plot shows electric fields at the cathode (blue line) and the center of the second cell (red line) from time domain simulation of CUPID photogun with input power envelope (orange shaded area) at \SI{275}{\mega\watt} peak power. The inset shows that the cathode reaches \SI{500}{\mega\volt/\metre} peak field whereas the second cell reaches \SI{490}{\mega\volt/\metre} peak field. Field balance to within 2\% is achieved between both cells.}
\end{figure}

Time domain simulations of the CUPID photogun were performed to study its behavior in the short-pulse regime. This simulation is crucial as it allows us to study the build-up of fields and their field balance.
Figure \ref{fig:pulse_time_domain} shows the normalized compressed pulse shape used for our simulations, calculated from rf pulse compressor built in \cite{Dhar2025} using formulae from \cite{nantista_radiofrequency_1995}. Figure \ref{fig:field_pulse}(a) shows the time domain simulation of peak fields on the cathode (blue line) at \SI{500}{\mega\volt/\metre} and the center of second cell (red line) at \SI{490}{\mega\volt/\metre} with an input rf pulse (orange shaded area) at \SI{275}{\mega\watt} peak power. Fields within the CUPID photogun reach their peak near steady-state in the time interval between \SI{256}{\nano\second} and \SI{257}{\nano\second} of the input pulse (inset of Fig.~\ref{fig:field_pulse}(a)), indicating that this is the time window where photoemission should happen such that electrons are accelerated with the highest field gradient. The \SI{275}{\mega\watt} peak power requirement is lower than the reported \SI{317}{\mega\watt} experiment in \cite{dhar:ipac2025-wepb100}.

\begin{figure}[h!]
\includegraphics[width=\linewidth]{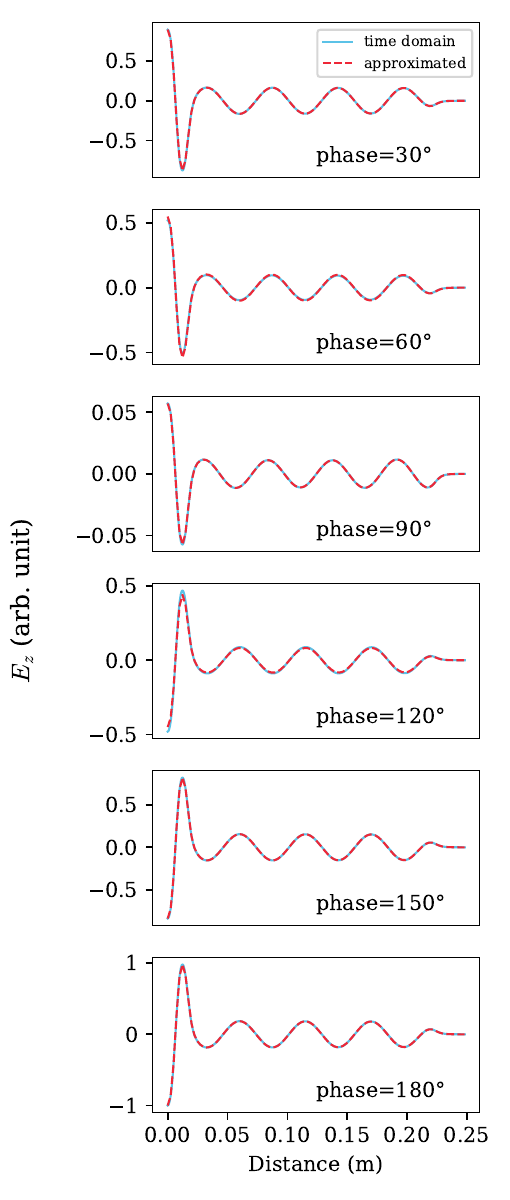}
\caption{\label{fig:field_pulse_compare}Plots of on-axis electric fields at different rf phases obtained using time domain simulations (blue lines) and Eq.~\ref{eq:traveling} (red lines). Both fields reach a good agreement and hence justify the use of approximated electric field for beam dynamics simulations. }
\end{figure}

\subsection{Rf pulse compression}

We intend to use rf pulse compression to generate these short-pulses necessary for CUPID, which provides a practical path to higher peak power at the cost of pulse length. One type of rf pulse compressor, the so-called SLED (SLAC Energy Development), uses low loss cavities fed by external waveguides to store energy~\cite{SLED}. Compression gain of the SLED is determined by the cavity's external quality factor, and the charging time. During this charging time, the cavity is filled with rf energy from the klystron. This stored energy is then compressed and rapidly dumped by flipping the phase of the incoming rf pulse. The speed of this phase flip ($t_{\mathrm{flip}}$) determines the compression gain of the SLED. Several types of these SLEDs have been developed at SLAC over the years~\cite{tantawi,frazi,wang,Dolgashev:2021jcv}. 

\begin{figure}[htb]
    \centering
	\includegraphics[clip=true,trim={0 0 0 7},width=\linewidth]{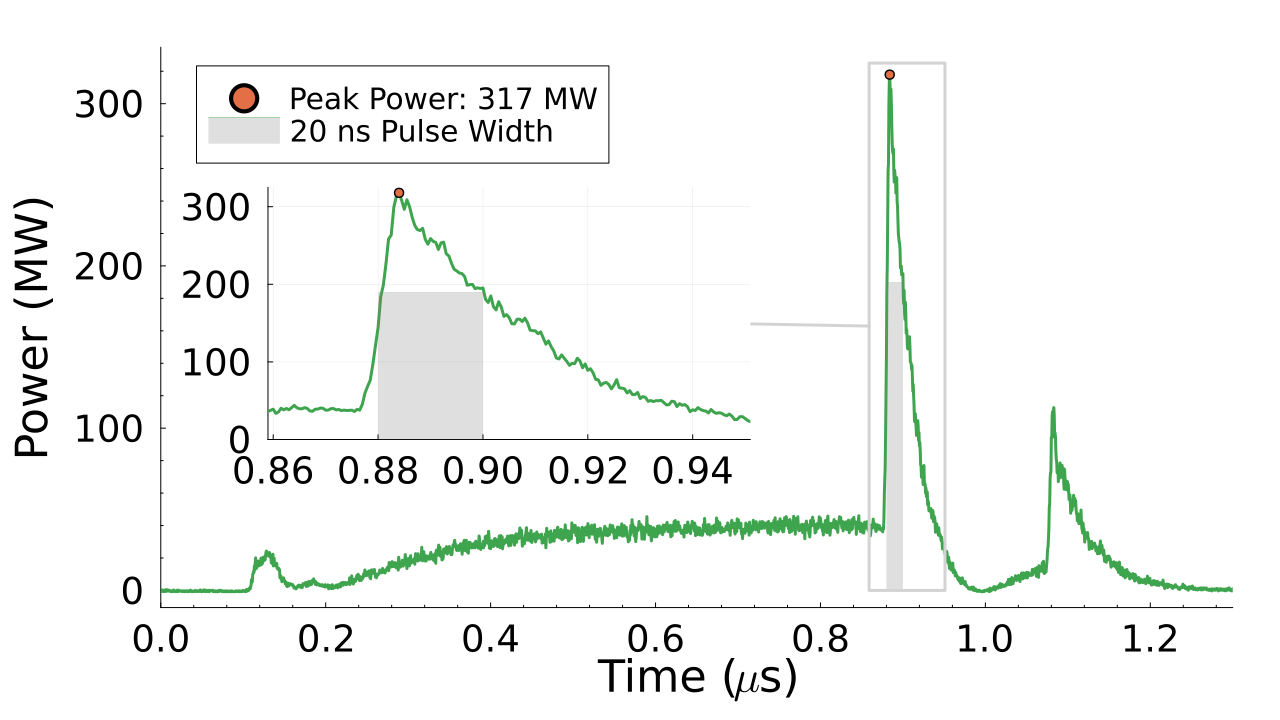}
    \caption{Measurement of the output SLED pulse. The forward power was \SI{52}{\mega\watt}, and the peak compressed power (green) is \SI{317}{\mega\watt}, yielding a compression factor of 6.1.}
    \label{fig:pulse}
\end{figure}

Most recently, a new \SI{11.424}{\giga\hertz} SLED was developed using two over-moded spherical cavities, each with an axially-symmetric TE$_{023}$ spherical cavity mode. This mode has no electric fields on the surface, improving the cavity's high power performance. This new SLED consists of two spherical cavities and a previously designed waveguide hybrid with circular ports. These ports operate with the TE$_{01}$ circular waveguide mode. Recent high power tests performed at SLAC have demonstrated peak rf power up to \SI{317}{\mega\watt} when powered by a single \SI{50}{\mega\watt} klystron~\cite{dhar:ipac2025-wepb100}. An example of this amplitude and phase data is shown in Fig.~\ref{fig:pulse}. This SLED demonstrated a compression factor of 6.1, and a Full-width Half-Maximum (FWHM) of \SI{20}{\nano\second}.

\section{\label{sec:cupid_beam_dynamics}CUPID Photoinjector Beam Dynamics}
\subsection{CUPID Photoinjector}
\begin{figure}[h!]
\includegraphics[width=\linewidth]{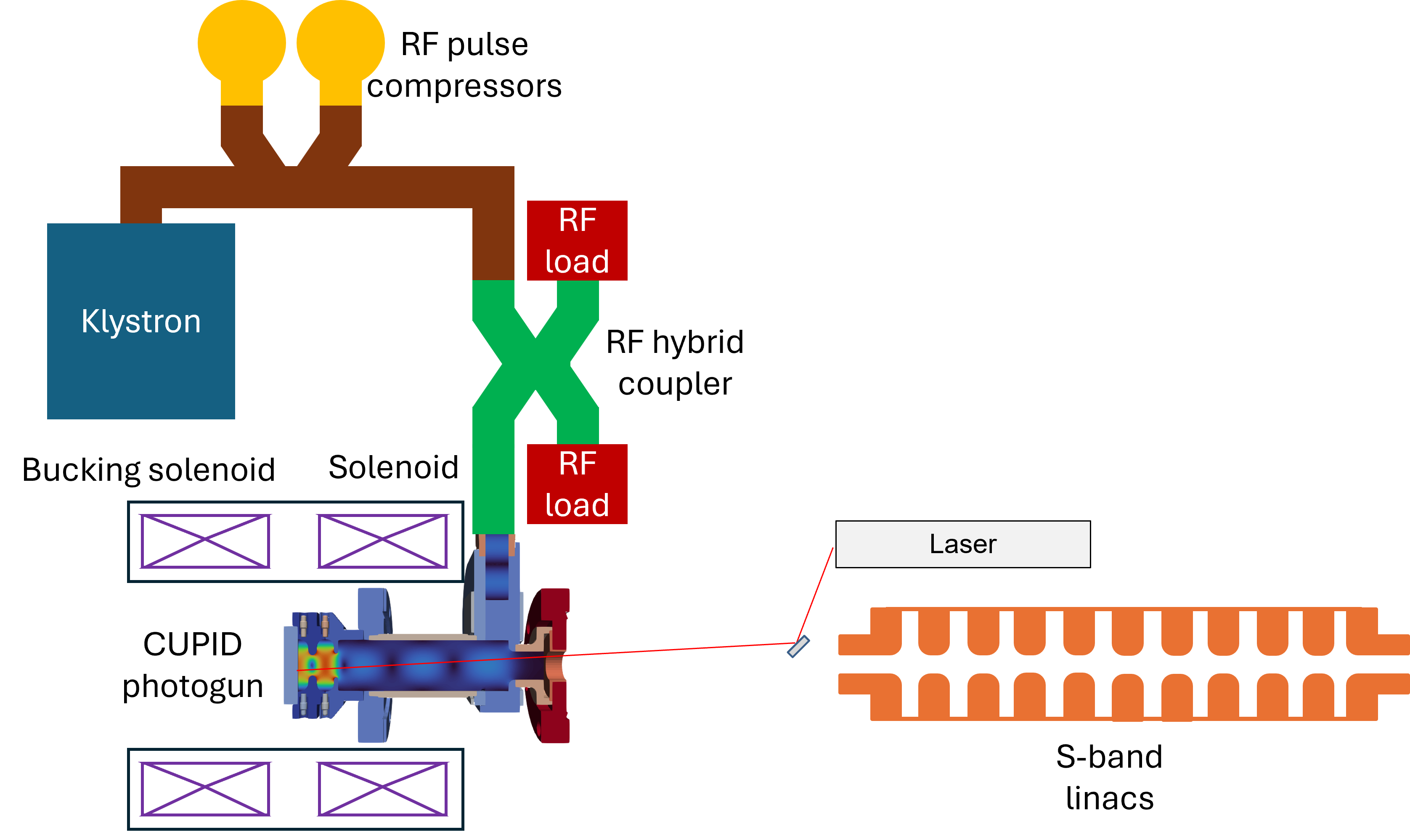}
\caption{\label{fig:cupid_photoinjector}Rf pulses generated from the klystron (blue box) is compressed by rf pulse compressors (yellow circles) into high power short rf pulses, which then sent to the rf hybrid coupler. Rf pulses are split into two output ports, where one of them is connected to our CUPID photogun. The other output port is connected to a rf load. Since more than 95 \% of rf power will be reflected from CUPID photogun, another rf load is connected to the last port of the hybrid coupler to prevent reflected pulses from damaging the klystron. CUPID photogun is enclosed by a superconducting solenoid and connected to downstream S-band linacs. It is noted that the use of rf hybrid coupler and rf loads is to prevent the damage of klystron by reflected rf pulses, which is due to our design choice of using an existing mode launcher with a single port. Alternative designs such as powering the photogun with two input ports will be investigated in the future.}
\end{figure}

The CUPID photoinjector consists of the CUPID photogun, a superconducting solenoid and three downstream \SI{1.2}{\metre} long \SI{2.856}{\giga\hertz} accelerating structures. Figure \ref{fig:cupid_photoinjector} shows the schematic diagram of the photoinjector. Rf pulses generated from the klystron (blue box) is compressed by rf pulse compressors (yellow circles) into high power short rf pulses, which then sent to the rf hybrid coupler. Rf pulses are split into two output ports, where one of them is connected to our CUPID photogun. The other output port is connected to a rf load. Since more than 95 \% of rf power will be reflected from CUPID photogun, another rf load is connected to the last port of the hybrid coupler to prevent reflected pulses from damaging the klystron. CUPID photogun is enclosed by a superconducting solenoid and connected to downstream S-band linacs. It is noted that the use of rf hybrid coupler and rf loads is to prevent the damage of klystron by reflected rf pulses, which is due to our design choice of using an existing mode launcher with a single port. Alternative designs such as powering the photogun with two input ports will be investigated in the future. 
\subsubsection{CUPID Photogun}
For an photogun operating in a steady-state regime, its on-axis field map $E_z(z,t)$ is described as
\begin{align}\label{eq:traveling}
    E_z(z, t)&=\Re{\{(E_{z,\text{real}}(z) + \mathrm{i} E_{z,\text{imag}}(z))\mathrm{e}^{\mathrm{i}\omega t}\}}\;,
\end{align}
where $E_{z,\text{real}}$ and $E_{z,\text{imag}}$ are the real and imaginary parts of an eigenmode solution of Maxwell equations, and 
$\Re\{\}$ is the operation of taking real part from the operand inside. We use this approach to approximate the on-axis field map of CUPID photogun for our subsequent beam dynamics studies using General Particle Tracer (GPT) simulation software \cite{generalparticletracer,vanderGeer:2003zb}, where the complex field maps were obtained from a snapshot in time of the time domain simulation shown in Fig.~\ref{fig:field_pulse}, at \SI{256}{\nano\second} where the field reaches quasi steady-state. We justify this approach by comparing it with time domain field maps taken from different snapshots in time, similar to the study in \cite{chen2025shortpulsedrivenradiofrequencyxband}. Figure \ref{fig:field_pulse_compare} shows the comparison at different rf phases and both fields reach a good agreement.

\begin{figure}[h!]
\centering
\includegraphics{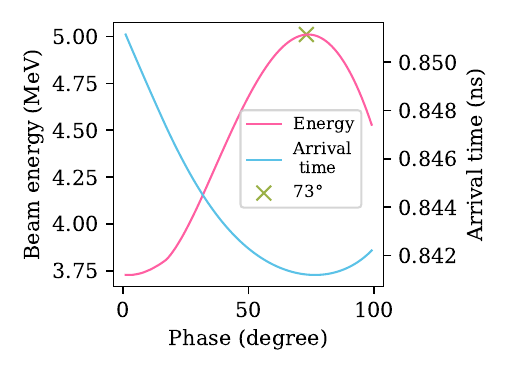}
\caption{\label{fig:phase_scan}Plots of beam energy (pink line) and time of arrival (blue line) at the exit of CUPID photogun (\SI{0.25}{\metre}) from the cathode at different rf phases, where the zero phase is defined as the phase where photoemission is first observed. The highest beam energy is \SI{5.01}{\mega\electronvolt} with the on-crest launch phase at \SI{73}{\degree} and the time of arrival less than \SI{0.9}{\nano\second}.}
\end{figure}

\begin{figure}[h!]
\includegraphics[width=\linewidth]{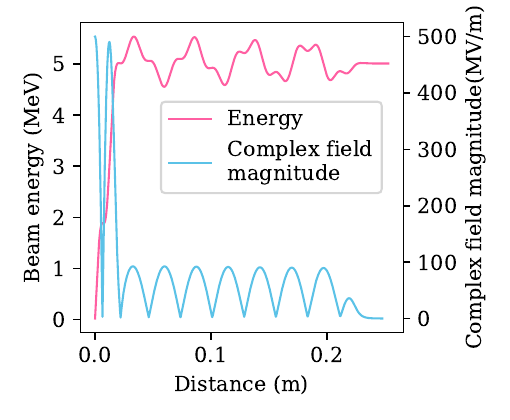}
\caption{\label{fig:energy_gain}The evolution of the beam energy versus distance (pink line) with the complex field magnitude (blue line) of CUPID photogun. As the phase of traveling waves is not adjusted to synchronize with the beam's phase, the beam experiences oscillation in its energy.}
\end{figure}

Figure \ref{fig:phase_scan} shows the beam energy (pink line) and time of arrival (blue line) at the exit of CUPID photogun (\SI{0.25}{\metre} from the cathode) with different rf phases, where the zero phase is defined as the phase where photoemission is first observed. The highest beam energy is \SI{5.01}{\mega\electronvolt} with the on-crest launch phase at \SI{73}{\degree} and the time of arrival less than \SI{0.9}{\nano\second}. The evolution of the beam energy versus distance is shown in Fig.~\ref{fig:energy_gain} (pink line) with the complex field magnitude (blue line) of CUPID photogun. Since CUPID photogun is designed to be over-coupled, there are non-negligible residual traveling waves present in the waveguide. As the phase of traveling waves is not adjusted to synchronize with the beam's phase, the beam experiences oscillation in its energy. Similar behavior was reported in ref.\cite{giribono_dynamics_2023}. We made this design choice for having a long enough waveguide coupler to place a magnetic solenoid close to the CUPID photogun. Alternatives will be investigated in the future \cite{gonzalez-iglesias_x-band_2021,faillace_high_2022}.

\subsubsection{Solenoid}
\begin{figure}[h!]
\includegraphics[width=\linewidth]{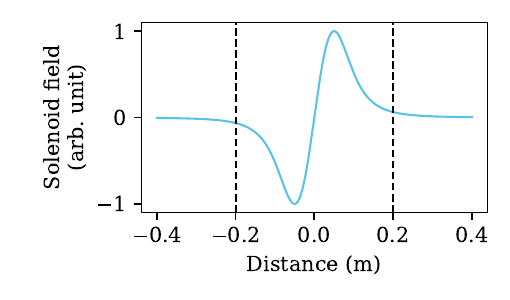}
\caption{\label{fig:sol}On-axis magnetic field profile of a superconducting solenoid used for the CUPID photoinjector, where the origin is the location of the cathode of the CUPID photogun. Black dashed lines are the exits of the solenoid. The modeling and design of this superconducting solenoid was provided by Cryomagnetics, Inc \cite{cryomagnetics}.}
\end{figure}

It is crucial to put a magnetic solenoid very close to an photogun to focus the photo-emitted electron beam. To provide a stable and high-strength magnetic field, we decided to use a superconducting solenoid for our CUPID photoinjector. It is envisioned that this superconducting solenoid will have a \SI{0.09}{\metre} bore diameter, \SI{0.4}{\metre} full length and provide up to \SI{1}{T} peak magnetic field. A bucking solenoid will be employed to cancel the magnetic field at the center of the solenoid. Figure \ref{fig:sol} shows the on-axis field profile of the superconducting solenoid, where the origin is the location of the cathode of CUPID photogun. Black dashed lines are the exits of the solenoid. The modeling and design of this superconducting solenoid was provided by Cryomagnetics, Inc \cite{cryomagnetics}.

\subsubsection{S-band Linacs}
\begin{figure}[h!]
\centering
\includegraphics[width=\linewidth]{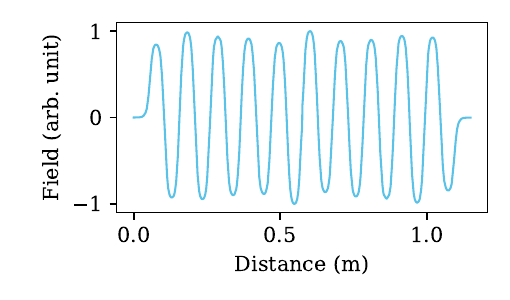}
\caption{\label{fig:eic} On-axis electric field of the \SI{1.2}{\metre} long S-band linacs operating at \SI{2.856}{\giga\hertz} are used for our CUPID photoinjector. Three such structures were used for beam dynamics simulations.}
\end{figure}

Three \SI{1.2}{\metre} long S-band linacs operating at \SI{2.856}{\giga\hertz} are used for our CUPID photoinjector. These linacs use the distributed-coupling design, cryogenic cooled, with each structure having 20 accelerating cells to provide on-axis peak field at \SI{100}{\mega\volt/\metre} \cite{Dhar:2024hgn}. Its on-axis field profile is shown in Fig.~\ref{fig:eic}. These structures have an aperture radius of \SI{14.12}{\milli\metre}, larger than that of the S-band linac in use at LCLS. This allows the structure to maintain beam emittance more easily. Due to the standing wave nature of the distributed coupling design, the accelerating fields are more stable as well.

\subsection{Beam Dynamics}
We performed beam dynamics simulation studies of the CUPID photoinjector using GPT.
Cylindrically-symmetric field maps were used for all accelerator components and three-dimensional mesh-based space-charge algorithm was used for \num{500000} macroparticles in our simulations.

We were interested in the generation of high brightness beams at \SI{100}{\pico\coulomb} total charge.
The initial beam distribution had a mean transverse energy (MTE) of \SI{400}{\milli\electronvolt}, which corresponds to the thermal emittance of \SI{0.9}{\micro\metre/\milli\metre}, given by
\begin{align}
    \frac{\varepsilon_n}{\sigma_x}=\sqrt{\frac{\text{MTE}}{mc^2}}\;,
\end{align}
where $\sigma_x$ is the laser spot size \cite{dowell_quantum_2009}. The beam was assumed to have a uniform radial and flat top temporal distribution. We chose this value of MTE to match the measured MTE of the current LCLS photoinjector, where a copper backplate is used as a cathode \cite{snowmass2021_fuhao,uk_seminar_raubenheimer}.
Lowering MTE can improve the beam brightness as the smallest achievable emittance is limited by the beam's MTE. Recent studies showed that the use of cryogenically-cooled cathode in conjunction with a well-ordered crystalline copper surface can minimize the MTE down to $\sim$\SI{5}{\milli\electronvolt} \cite{PhysRevLett.125.054801}. However, low MTE comes at the expense of low amount photoemitted electrons, quantified by the quantum efficiency (QE) for the same input laser power. While increasing the laser power could solve the problem, the beam's MTE might be increased due to the nonlinear photoemission effects of a high power laser, as demonstrated in \cite{PhysRevAccelBeams.26.093401}. In addition to that, surface roughness will further increase MTE due to the presence of high fields \cite{PhysRevAccelBeams.21.093401}. The use of alkali-antimonide thin films can provide high QE and low MTE to mitigate these issues \cite{Pallavi101063}. Since our paper focuses on the improvement of beam brightness brought by the high gradient photogun, we chose \SI{400}{\milli\electronvolt} for subsequent simulations, where we assumed that this is a reasonable parameter based on the measured MTE of the current LCLS photoinjector.

\begin{figure}[h!]
\centering
\includegraphics[width=\linewidth]{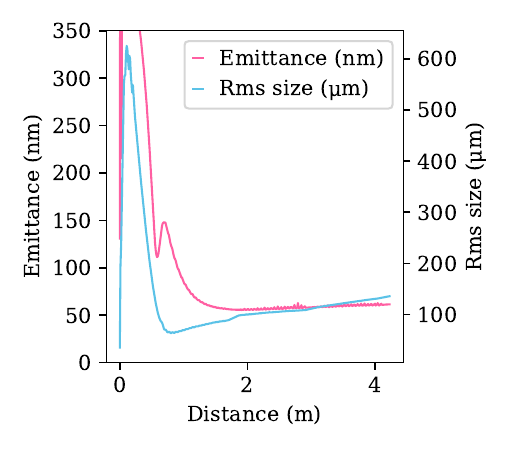}
\caption{\label{fig:enx}The evolution of beam's emittance (pink line) and rms beam size (blue line) of CUPID photoinjector using parameters shown in Tab.~\ref{tab:beam}.}
\end{figure}

\begin{figure}[h!]
\centering
\includegraphics[width=\linewidth]{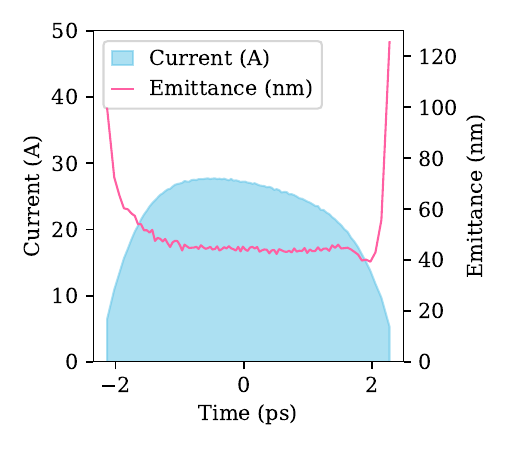}
\caption{\label{fig:enx_slice}Current profile (blue shaded region) and slice emittance (pink line) of the beam at the end of CUPID photoinjector. The core emittance (at center) is \SI{40}{\nano\metre}.}
\end{figure}

We employed the emittance compensation scheme to achieve low emittance beams by focusing the beam to its beam waist to realign different longitudinal slices of the beam, followed by accelerating the beam to a very high energy to freeze the phase space oscillation of different longitudinal beam slices \cite{CARLSTEN1989313,serafini-1997-a,Ferrario:2000nn}. Figure \ref{fig:enx} shows the evolution of electron beam's transverse emittance (pink line) and its rms size (blue line) along the CUPID photoinjector.
We achieved a \SI{63}{\nano\metre} beam emittance using the CUPID photoinjector. Figure \ref{fig:enx_slice} shows the current profile (blue shaded area) and the beam's slice emittance (pink line). The beam core has \SI{40}{\nano\metre} slice emittance and \SI{28}{\ampere} peak current.
We note that CUPID photoinjector needs only one accelerating structure for emittance compensation, as shown in Fig.~\ref{fig:eic} where the beam's emittance reaches a minimum near a distance of \SI{2}{\metre}. Two additional structures were included in our studies to match the output beam energy of the current LCLS injector at \SI{135}{\mega\electronvolt} \cite{PhysRevSTAB.11.030703}. We summarize our CUPID photoinjector and output beam parameters in Tab.~\ref{tab:beam}.

\begin{figure}[h!]
\centering
\includegraphics[width=\linewidth]{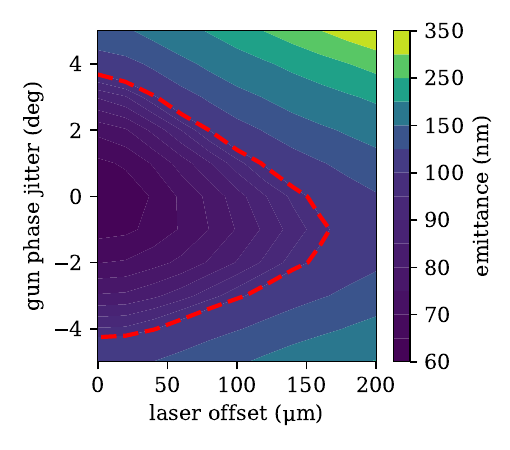}
\caption{\label{fig:enx_jitter}Contour plot of beam's emittance versus the CUPID gun phase jitter and laser offset. The region of tolerance with less than \SI{100}{\nano\metre} emittance is covered by a red dashed line. This region indicates the required gun phase jitter and laser offset to achieve less than \SI{100}{\nano\metre} emittance out of the CUPID photoinjector.}
\end{figure}

We performed a tolerance study where we varied CUPID gun phase jitter and input laser offset. Figure \ref{fig:enx_jitter} shows the result in the form of a contour plot of beam's emittance versus gun phase jitter and laser offset. Using \SI{100}{\nano\metre} emittance as the tolerance level, we drew the region of acceptable gun phase jitter and laser offset with a red dashed-line in Fig.~\ref{fig:enx_jitter}. This result serves as a reference for us to develop low-level radiofrequency (llrf) systems for our klystrons and pulse compressors to achieve good phase stability \cite{liu2025highprecisionrfpulse,liu2025generationdirectrfsampling}.

Note that our beam dynamics simulations did not take into account intrabeam scattering (IBS). IBS occurs due to short-range interactions between electrons, increasing the beam's slice energy spread (SES) and degrading the six-dimensional brightness. IBS was studied in the past where it was concluded that the increase of IBS-induced SES was insufficient to affect the performance of x-ray FELs \cite{huang_intrabeam_2002}. In addition, a laser heater was introduced to increase the SES to suppress the microbunching instability, such as in LCLS.
Recent improvements in electron beam quality and SES measurement methods were able to confirm the effect of IBS in photoinjectors of modern FELs \cite{di_mitri_experimental_2020,PhysRevAccelBeams.23.090701,PhysRevAccelBeams.25.104401,qian_slice_2022}. IBS-induced SES could be significant for the \SI{100}{\pico\coulomb} low emittance beam that we were studying with CUPID, as reported in \cite{robles_versatile_2021}. However, accurate simulation of IBS requires the calculation of one-to-one Coulomb interaction and thus requires a huge computational cost.
Scaled-down simulations in ref \cite{robles_versatile_2021} with similar beam parameters showed that IBS yielded less than \SI{1}{\kilo\electronvolt} SES, which was still insufficient to affect the FEL performance that we are targeting. Experimental evidence of IBS-induced SES drove the theoretical and computational efforts in predicting IBS-induced SES, as reported in refs \cite{valdor:ipac2025-weps042,lucas2025accuratemodellingintrabeamscattering}.
On the other hand, IBS-induced SES can be used to suppress microbunching instability with some clever designs of accelerator lattice, as proposed in refs \cite{di_mitri_estimate_2014,di_mitri_experimental_2020,QIANG2023167968}. 

\begin{table}[b]
\caption{\label{tab:beam}Optimized Parameters for the Injector and Final Beam Parameters}
\begin{ruledtabular}
\begin{tabular}{lcc}
\textrm{Parameter}&
\textrm{Value}&
\textrm{Unit}\\
\colrule
Laser spot size & $70$ & \si{\micro\metre} \\
Laser duration & 4.5 & \si{\pico\second}\\
Mean transverse energy & \num{400} & \si{\milli\electronvolt}\\
Beam charge & 100 & \si{\pico\coulomb}\\
Gun frequency& \num{11.424} & \si{\giga\hertz} \\
Field on cathode & \num{500} &  \si{\mega\volt/\metre}\\
Injection phase & \num{70} &  \si{\degree}\\
Linacs frequency& \num{2.856} & \si{\giga\hertz} \\
Linac 1 gradient & \num{100} &  \si{\mega\volt/\metre} \\
Linac 1 phase (from on-crest) & \num{-3.3} &  \si{\degree}\\
Linac 1 distance from the cathode & 0.6 & \si{\metre}\\
Linac 2 gradient & \num{100} &  \si{\mega\volt/\metre} \\
Linac 2 phase (from on-crest) & \num{0} &  \si{\degree}\\
Linac 3 gradient & \num{44} &  \si{\mega\volt/\metre} \\
Linac 3 phase (from on-crest) & \num{0} &  \si{\degree}\\
Solenoid field & 0.62 & \si{\tesla}\\
Solenoid peak from the cathode & 0.05 & \si{\metre}\\
\colrule
Final beam energy & \num{135} & \si{\mega\electronvolt} \\
Final beam bunch length & \num{342} & \si{\micro\metre} \\
Final beam rms size & \si{136} & \si{\micro\metre} \\
Final beam transverse emittance & \num{63} & \si{\nano\metre} \\
Final beam relative energy spread & \num{0.04} & \%\\
\end{tabular}
\end{ruledtabular}
\end{table}

\section{\label{sec:cupid_lcls}LCLS with CUPID}
\begin{figure*}
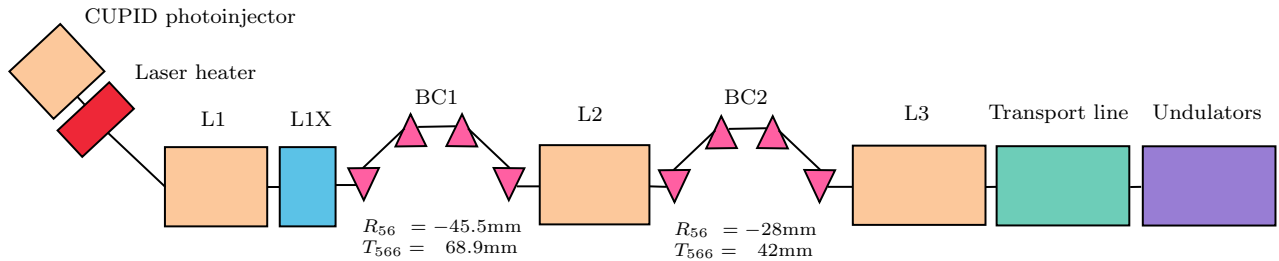

\include{lcls_layout}
\caption{\label{fig:fig_lcls}Schematic diagram of the LCLS copper accelerator with CUPID photoinjector and a laser heater. LCLS copper accelerator consists of three linac sections L1, L2 and L3, an X-band harmonic linearizer section L1X, two bunch compressors BC1 and BC2, a long transport beamline to deliver beams to the x-ray undulator beamline.}
\end{figure*}
With the low emittance beams obtained from CUPID photoinjector, we investigated the use of low emittance beams in the current LCLS copper accelerator to examine the potential performance improvement of the x-ray FEL. The LCLS copper accelerator consists of three linac sections L1, L2 and L3, an X-band harmonic linearizer section, two bunch compressors BC1 and BC2, a long transport beamline and a beam switchyard to deliver beams to the x-ray undulator beamline. Figure \ref{fig:fig_lcls} shows the schematic diagram of the LCLS copper accelerator with the CUPID photoinjector.

At present, LCLS first accelerates and manipulates the electron beam's longitudinal phase space at L1 and the X-band linearizer section, then it compresses the beam, followed by another acceleration and compression at L2 and BC2, further acceleration at L3, and finally delivers the beam through the transport line and beam switchyard to the x-ray undulator beamline for coherent x-ray generation. The source of beam degradation in the LCLS copper acclerator includes wakefields in linac sections, coherent synchrotron radiation in bend sections (BC1, BC2 and beam switchyard), and longitudinal space charge throughout the whole accelerator. As the beam brightness is determined at its source, suppressing emittance growth is an important topic for a FEL.

In this section, we present our beam dynamics studies of LCLS with the CUPID photoinjector. In particular, we were interested in the final beam parameter at the beginning of the x-ray undulator beamline for an energy range $\SI{14}{\giga\electronvolt}\sim\SI{15}{\giga\electronvolt}$ for very hard x-ray generation and high pulse energy. 

\subsection{Simulation Setup for LCLS Copper Accelerator}
The initial beam distribution used for LCLS beam dynamics studies was obtained from the output of the CUPID photoinjector in the previous section. We used the simulation software \verb|elegant| version 2025.2 for our studies \cite{borland_elegant:_2000}. The output beam distribution file was converted from GPT format to an \verb|elegant|-compatible format. In addition, we resampled the beam distribution using \verb|elegant| script \verb|smoothDist6s| to generate low-noise distribution to suppress numerical noise. We also transformed the transverse phase space to fit the Twiss parameters of the LCLS copper accelerator entrance. Finally, we used a laser heater to increase the beam's SES by a few \si{\kilo\electronvolt} to suppress microbunching instability. It is noted that the effect of IBS could provide an increase in SES and thus eliminate the need of a dedicated laser heater. We reserve the topic of IBS and microbunching instability for our future studies.

\subsection{LCLS Copper Accelerator Beam Dynamics with CUPID}
\begin{figure}[h!]
\centering
\includegraphics[width=\linewidth]{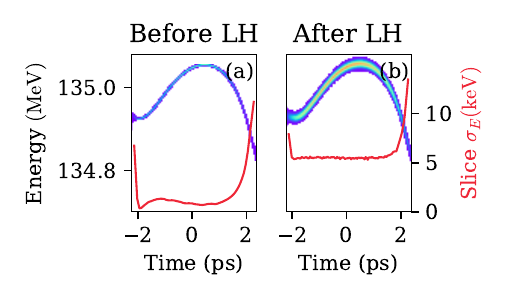}
\caption{\label{fig:lps_heater}Longitudinal phase spaces of the beam (a) before laser heater and (b) after laser heater. Slice energy spread (red line) increases by \SI{5}{\kilo\electronvolt} after the laser heater.}
\end{figure}

\begin{figure*}
\includegraphics{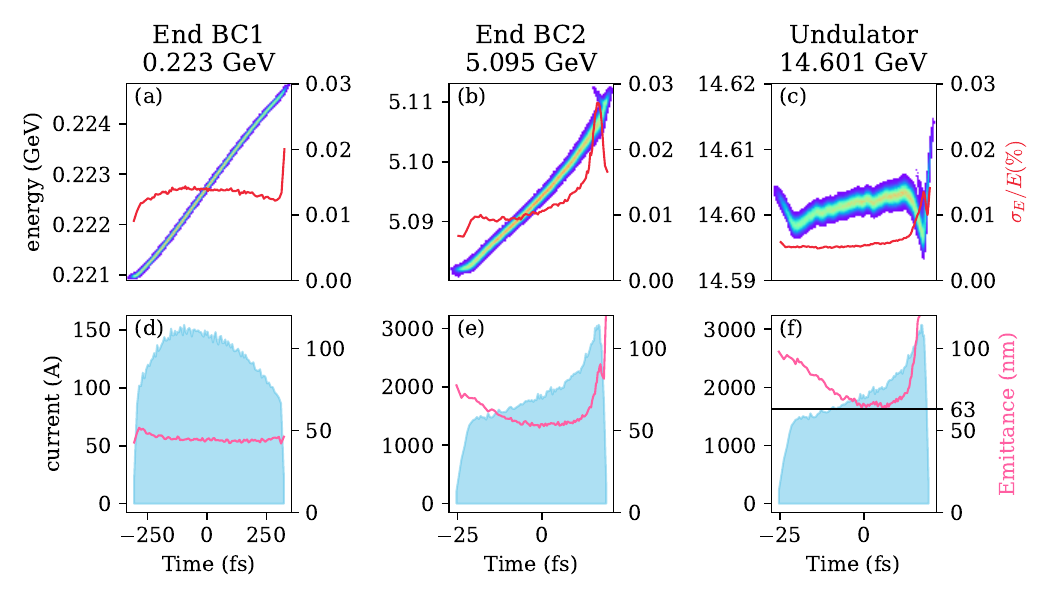}
\caption{\label{fig:lps_lcls_beam}Longitudinal phase space and current profile respectively at (a)(d) the end of BC1, (b)(e) the end of BC2 and (c)(f) the entrance of undulator beamline. The beam was compressed and collimated at BC1 to remove current horns on both ends of the beam. It was further compressed at BC2 to reach \SI{1600}{\ampere} average current. The beam was further accelerated and transported to the entrance of undulator beamline at \SI{14.601}{\giga\electronvolt}. Slice emittances were increased significantly but overall remained less than \SI{100}{\nano\metre}.}
\end{figure*}

Here we present LCLS beam dynamics simulation results using \verb|elegant|. The longitudinal phase space (LPS) of the initial beam distribution is shown in Fig.~\ref{fig:lps_heater} where the beam's SES was increased by \SI{5}{\kilo\electronvolt} to suppress microbunching instability. The beam distribution was then transported through the whole LCLS copper accelerator using parameters shown in Tab.~\ref{tab:lcls_beam}. Figure \ref{fig:lps_lcls_beam} shows the LPS and the current profile respectively at (a)(d) the end of BC1, (b)(e) the end of BC2 and (c)(f) the entrance of undulator beamline.
As the bending angles of bunch compressors BC1 and BC2 are fixed, we tuned linac phases to adjust compression factor.
Rf phases of linac section L1 and X-band linearizer section L1X were adjusted to compress the beam to about \SI{150}{\ampere} peak current. A collimator in the mid section of BC1 were used to collimate the head and tail part of the beam to suppress current horns. The beam's total charge was reduced to \SI{81}{\pico\coulomb} after passing through the collimator.

The beam was further accelerated through linac section L2 and compressed at BC2 to reach a peak current at \SI{3000}{\ampere} and \SI{1600}{\ampere} at the core of the beam, as shown in Fig.~\ref{fig:lps_lcls_beam}(b). The beam's slice emittances were increased after passing through BC2 as shown in Fig.~\ref{fig:lps_lcls_beam}(e), where the slice emittances of the both ends of beam were increased significantly compared to other beam slices. The beam reached final energy at \SI{14.601}{\giga\electronvolt} at the entrance of undulator beamline as shown in Fig.~\ref{fig:lps_lcls_beam}(c). Slice emittances were increased significantly at both ends of beam but overall less than \SI{100}{\nano\metre} as shown in Fig.~\ref{fig:lps_lcls_beam}(f). The beam core (at \SI{0}{\femto\second}) had \SI{63}{\nano\metre} slice emittance and \SI{0.005}{\percent} relative energy spread.

\begin{figure}[h!]
\centering
\includegraphics[width=\linewidth]{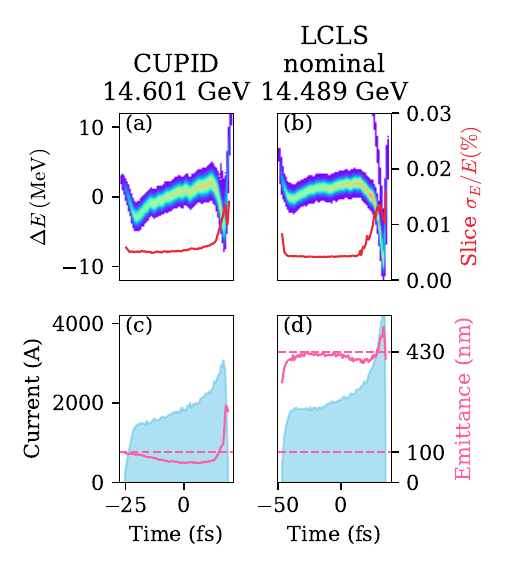}
\caption{\label{fig:lps_lcls_compare}Comparison of longitudinal phase spaces using (a)(c) CUPID photoinjector and (b)(d) LCLS nominal injector. LCLS nominal beam has a higher average current at \SI{2000}{\ampere} than CUPID beam at \SI{1600}{\ampere} but with much larger slice emittance at \SI{430}{\nano\metre} (dashed line at (d)), whereas CUPID beam has overall less than \SI{100}{\nano\metre} slice emittance (dashed line at (c)). Both have comparable slice energy spread (red lines at (a)(b)).}
\end{figure}

We performed simulations with \SI{250}{\pico\coulomb} nominal beam output from the existing LCLS injector and compare its results with the CUPID photoinjector. We hereby refer to this case as ``LCLS nominal". Accelerator parameters for LCLS nominal were similar to that of CUPID with minor changes in rf phases.
Similarly, LCLS nominal beam was collimated at BC1 and reduced to \SI{184}{\pico\coulomb} charge to suppress current horns. Figure \ref{fig:lps_lcls_compare} shows LPS and current profiles at the undulator entrance using (a)(c) CUPID photoinjector and (b)(d) LCLS nominal.
LCLS nominal beam has a higher average current at \SI{2000}{\ampere} than CUPID beam at \SI{1600}{\ampere} but with much larger slice emittance at \SI{430}{\nano\metre} (dashed line at (d)), whereas CUPID beam has overall less than \SI{100}{\nano\metre} slice emittance (dashed line at (c)). Both have comparable slice energy spread.

Our results show a promising outcome with merely swapping the existing LCLS injector with CUPID photoinjector.
Increases of the beam's emittance were caused mainly by coherent synchrotron radiation effects at bunch compressors BC1, BC2 and the dispersive section of the long transport beamline. It is noted that the suppression of current horns on both ends of beam can be achieved using octupole magnets on both bunch compressors instead of collimating the beam \cite{sudar_octupole_2020}. Similarly, IBS-induced SES could be used instead of a laser heater to suppress microbunching instability \cite{QIANG2023167968}. In our studies, we chose to stick closely with the existing LCLS copper accelerator configuration to demonstrate the improvement of beam quality from the CUPID photoinjector. Additional upgrades and improvements to the LCLS copper accelerator complex would further enhance the benefits possible with a CUPID upgrade.
\begin{table}[b]
\caption{\label{tab:lcls_beam}Simulation Parameters for the LCLS copper accelerator and Final Beam Parameters}
\begin{ruledtabular}
\begin{tabular}{lcc}
\textrm{Parameter}&
\textrm{Value}&
\textrm{Unit}\\
\colrule
L1 phase (from on-crest) & \num{-18} &  \si{\degree}\\
L1X phase (from on-crest) & \num{-155.74} & \si{\degree}\\
L2 phase (from on-crest) & \num{-31.7} &  \si{\degree}\\
L3 phase (from on-crest) & \num{0} &  \si{\degree}\\
BC1 $R_{56}$ & \num{-45.5} & \si{\milli\metre}\\
BC1 $T_{566}$ & \num{68.9} & \si{\milli\metre}\\
BC2 $R_{56}$ & \num{-28} & \si{\milli\metre}\\
BC2 $T_{566}$ & \num{42} & \si{\milli\metre}\\
\colrule
Total charge (after collimator) & \num{81} & \si{\pico\coulomb}\\
Beam energy (initial) & \num{135} & \si{\mega\electronvolt} \\
Beam energy (after BC1) & \num{223} & \si{\mega\electronvolt} \\
Beam energy (after BC2) & \num{5.095} & \si{\giga\electronvolt} \\
Beam energy (undulator entrance) & \num{14.601} & \si{\giga\electronvolt} \\
Core emittance (undulator entrance) & \num{63} & \si{\nano\metre} \\
Core relative energy spread & \num{0.005} & \%\\
\end{tabular}
\end{ruledtabular}
\end{table}

\subsection{Simulation Setup for FEL}
Beams from \verb|elegant| simulations for both CUPID and LCLS nominal were transformed to match the Twiss parameters of the entrance of LCLS hard x-ray undulator beamline. This undulator beamline is \SI{150}{\metre} long, with undulator period $\lambda_u=\SI{26}{\milli\metre}$ and variable gap for adjustable undulator parameter $K_u$ \cite{Leitner:IPAC2017-TUPAB123}. We used software \verb|GENESIS1.3| \cite{REICHE1999243} version 4.6.9 for our subsequent FEL simulations at photon energies \num{20}, \num{30}, \num{40}, \num{50} and \SI{60}{\kilo\electronvolt} to demonstrate performance improvement of x-ray generation from CUPID. Throughout our FEL simulations, undulator parameter $K_u$ were adjusted to match the FEL resonance condition, using
\begin{align}
    \lambda_\text{ph} &=\frac{\lambda_u}{2\gamma^2}\left(1 + \frac{{K_u}^2}{2}\right)\;,
\end{align}
where $\lambda_\text{ph}$ is the wavelength of x-ray photon, $\lambda_u$ is the undulator period, $\gamma$ is the beam's Lorentz factor, $K_u$ is the undulator parameter.

\subsection{LCLS FEL Performance with CUPID}
\begin{figure}[h!]
\centering
\includegraphics[width=\linewidth]{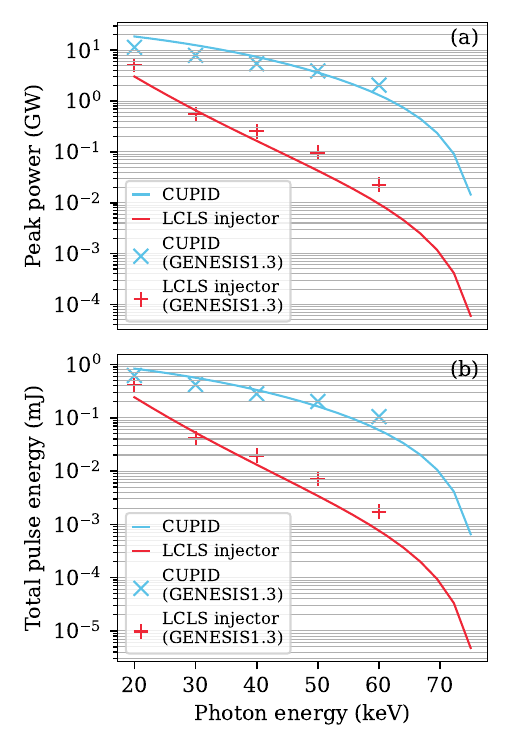}
\caption{\label{fig:fel_compare}Comparison of hard x-ray FEL performance at different photon energies in terms of (a) peak power and (b) total pulse energy. Ming-Xie parameterization with beams from Fig.~\ref{fig:lps_lcls_compare} were used. CUPID (blue line) outperforms LCLS nominal (red line) in both power and total pulse energy. \texttt{GENESIS1.3} simulation results show a good agreement (``$\times$" sign for CUPID and ``+" sign for LCLS nominal). CUPID maintains more than \SI{0.1}{\milli\joule} pulse energy up to \SI{60}{\kilo\electronvolt} photon energy.}
\end{figure}
\begin{figure}[h!]
\centering
\includegraphics[width=\linewidth]{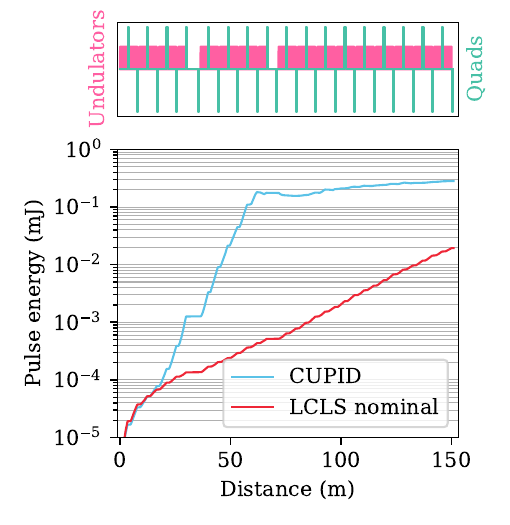}
\caption{\label{fig:fel_gain}Comparison of total pulse energy for \SI{40}{\kilo\electronvolt} x-ray photons using CUPID (blue line) and LCLS nominal beams (red line) with undulator beamline layout on top. The present undulator beamline is shorter than the saturation length required for LCLS nominal at \SI{40}{\kilo\electronvolt}, whereas CUPID reaches saturation after \SI{50}{\metre}.}
\end{figure}
We summarize the FEL saturated pulse energy and power as functions of photon energy in Fig.~\ref{fig:fel_compare} for CUPID (``$\times$" sign in Fig.~\ref{fig:fel_compare}) and LCLS nominal (``+" sign in Fig.~\ref{fig:fel_compare}). In addition to \verb|GENESIS1.3| simulations, we performed Ming-Xie parameterization analysis and its results (solid lines in Fig.~\ref{fig:fel_compare}) agree well with \verb|GENESIS1.3| simulations.
In general, CUPID performs better in all cases despite having lower total charge and average current.
The FEL performance difference between CUPID and LCLS nominal becomes significant as photon energy goes higher. Total pulse energy of LCLS nominal for \SI{40}{\kilo\electronvolt} photons drops to below \SI{0.02}{\milli\joule}, whereas CUPID maintains a pulse energy of \SI{0.32}{\milli\joule}.
We show an example of FEL gain profiles at \SI{40}{\kilo\electronvolt} for CUPID and LCLS nominal in Fig.~\ref{fig:fel_gain}. This is partly due to the fact that the present undulator beamline is shorter than the saturation length required for LCLS nominal to produce \SI{40}{\kilo\electronvolt} photons, whereas CUPID reaches saturation after \SI{50}{\metre}.
In overall, CUPID maintains pulse energies over \SI{0.1}{\milli\joule} for up to \SI{60}{\kilo\electronvolt} photons before reaching the \SI{150}{\metre} saturation length of the present LCLS undulator beamline. Ming Xie parameterization estimates that photon energy as high as \SI{70}{\kilo\electronvolt} with a pulse energy of \SI{0.01}{\milli\joule} is achievable. In addition, it is noted that the present LCLS undulator beamline was designed for a photon energy range of $\SI{1}{\kilo\electronvolt}\sim\SI{25}{\kilo\electronvolt}$. Undulators tailored for very hard x-ray generation would further improve the FEL performance.

\section{\label{sec:summary}Summary}
We have presented CUPID, a high gradient photogun capable of delivering high brightness electron beams down to \SI{63}{\nano\metre} transverse emittance at \SI{100}{\pico\coulomb}. It is achieved by forming a photoinjector consists of the CUPID photogun, a dedicated superconducting solenoid, and downstream S-band accelerating structures.
The CUPID photogun is designed to operate at \SI{500}{\mega\volt/\metre} field at the cathode driven by high power nanosecond rf pulses. SLAC-built klystrons and recently commissioned rf pulse compressors are used to drive the CUPID photogun, instead of the wakefield acceleration approach of AWA, to benefit from klystron phase stability and arbitrary rf pulse shapes \cite{liu2025highprecisionrfpulse,liu2025generationdirectrfsampling}.

Using the CUPID photoinjector for the existing LCLS copper accelerator, our \verb|elegant| simulations showed that final beam output for FEL is deliverable without modifications to the LCLS copper accelerator configuration with slice emittances of beam core below \SI{100}{\nano\metre}. This is four times improvement over the LCLS nominal beam at \SI{430}{\nano\metre} despite having lower average current. FEL simulations using \verb|GENESIS1.3| shows that CUPID beam delivers more than \SI{0.1}{\milli\joule} pulse energy up to \SI{60}{\kilo\electronvolt} using the present hard x-ray undulator beamline of LCLS.

Experimental efforts on CUPID are ongoing where CUPID prototypes of different copper alloys were fabricated for rf testing. In parallel, rf pulse compressors and X-band klystrons were commissioned and have demonstrated more than \SI{500}{\mega\watt} peak power, exceeding the requirements of the CUPID photogun. Near future plans include using rf pulse shaping for efficient peak power, and high power rf testing of CUPID photogun prototypes to investigate its feasibility. After that, the next step would be to commission the CUPID photogun in a dedicated accelerator test facility for electron beam generation. In addition, investigating the possibility of short pulse driven photogun at lower frequency bands would be an important topic for different accelerator applications. Investigating microscopic Coulomb interactions such as IBS would also be important.

CUPID represents a major step towards high gradient photoguns through the use of short pulses driven by existing klystron technology. Our work here shows that performance improvement of LCLS FEL made possible through the use of the CUPID photogun. Newer accelerator technologies could be incorporated to achieve even better performance. Synergy between these topics and CUPID will be subjects for future studies.

\begin{acknowledgments}
W.H.T thanks Zenghai Li, David Dowell, Valery Dolgashev, David Cesar and Emma Snively (SLAC), the Argonne Wakefield Accelerator group (ANL) for valuable discussions, R. Soliday (ANL), Ji Qiang (LBNL) and Andrea Latina (CERN) for help on simulations.
This work is supported by the U.S. Department of Energy Contract No. DE-AC02-76SF00515 with SLAC National Accelerator Laboratory.
Simulation works used resources of the National Energy Research Scientific Computing (NERSC) Center and the SLAC Shared Science Data Facility (S3DF) at SLAC National Accelerator Laboratory. NERSC is a U.S. Department of Energy Office of Science User Facility located at Lawrence Berkeley National Laboratory, operated under Contract No. DE-AC02-05CH11231.
\end{acknowledgments}


\bibliography{prab_cupid}

\end{document}

%% file: lcls_layout.tex
\tikzset{every picture/.style={line width=0.75pt}} 

\begin{tikzpicture}[x=0.75pt,y=0.75pt,yscale=-1,xscale=1]

\draw  [fill={rgb, 255:red, 252; green, 200; blue, 155 }  ,fill opacity=1 ] (87.4,109.87) -- (138.99,109.87) -- (138.99,149.87) -- (87.4,149.87) -- cycle ;
\draw  [fill={rgb, 255:red, 91; green, 194; blue, 231 }  ,fill opacity=1 ] (145.31,109.87) -- (173.47,109.87) -- (173.47,149.87) -- (145.31,149.87) -- cycle ;
\draw  [fill={rgb, 255:red, 255; green, 95; blue, 162 }  ,fill opacity=1 ] (187.7,136.67) -- (179.86,120.54) -- (195.35,120.39) -- cycle ;
\draw  [fill={rgb, 255:red, 255; green, 95; blue, 162 }  ,fill opacity=1 ] (211.29,93.24) -- (219.16,109.34) -- (203.68,109.54) -- cycle ;
\draw    (208.3,100.44) -- (197.77,110.44) -- (187.23,120.44) ;

\draw    (215.11,99.67) -- (234.07,99.33) ;
\draw  [fill={rgb, 255:red, 255; green, 95; blue, 162 }  ,fill opacity=1 ] (260.84,136.67) -- (268.68,120.54) -- (253.19,120.39) -- cycle ;
\draw  [fill={rgb, 255:red, 255; green, 95; blue, 162 }  ,fill opacity=1 ] (237.25,93.25) -- (229.37,109.34) -- (244.86,109.54) -- cycle ;
\draw    (240.24,100.44) -- (250.77,110.44) -- (261.31,120.44) ;

\draw  [fill={rgb, 255:red, 252; green, 200; blue, 155 }  ,fill opacity=1 ] (276.44,108.87) -- (331.75,108.87) -- (331.75,148.87) -- (276.44,148.87) -- cycle ;
\draw  [fill={rgb, 255:red, 252; green, 200; blue, 155 }  ,fill opacity=1 ] (434.33,109.12) -- (501.25,109.12) -- (501.25,149.12) -- (434.33,149.12) -- cycle ;
\draw  [fill={rgb, 255:red, 255; green, 95; blue, 162 }  ,fill opacity=1 ] (344.4,137.67) -- (336.56,121.54) -- (352.05,121.39) -- cycle ;
\draw  [fill={rgb, 255:red, 255; green, 95; blue, 162 }  ,fill opacity=1 ] (367.99,94.24) -- (375.86,110.34) -- (360.38,110.54) -- cycle ;
\draw    (365,101.44) -- (354.47,111.44) -- (343.93,121.44) ;

\draw    (371.81,100.67) -- (390.77,100.33) ;
\draw  [fill={rgb, 255:red, 255; green, 95; blue, 162 }  ,fill opacity=1 ] (417.54,137.67) -- (425.38,121.54) -- (409.89,121.39) -- cycle ;
\draw  [fill={rgb, 255:red, 255; green, 95; blue, 162 }  ,fill opacity=1 ] (393.95,94.25) -- (386.07,110.34) -- (401.56,110.54) -- cycle ;
\draw    (396.94,101.44) -- (407.47,111.44) -- (418.01,121.44) ;

\draw    (138.99,130) -- (145.31,130) ;
\draw    (173.49,129) -- (184.03,129) ;
\draw    (264.88,129.67) -- (276.47,129.67) ;
\draw    (331.91,129.67) -- (340.33,129.92) ;
\draw    (421.78,129.92) -- (434.35,129.92) ;
\draw  [fill={rgb, 255:red, 238; green, 39; blue, 55 }  ,fill opacity=1 ] (59.32,77.15) -- (71.76,90.49) -- (45.7,114.79) -- (33.26,101.45) -- cycle ;
\draw    (58.67,102.67) -- (87.64,130) ;
\draw  [fill={rgb, 255:red, 252; green, 200; blue, 155 }  ,fill opacity=1 ] (34.61,47.74) -- (56.9,71.64) -- (31.25,95.56) -- (8.96,71.66) -- cycle ;
\draw    (43.13,84.33) -- (46.82,88.33) ;
\draw  [fill={rgb, 255:red, 152; green, 125; blue, 212 }  ,fill opacity=1 ] (580.58,109.45) -- (647.5,109.45) -- (647.5,149.45) -- (580.58,149.45) -- cycle ;
\draw  [fill={rgb, 255:red, 109; green, 205; blue, 184 }  ,fill opacity=1 ] (506.86,109.4) -- (573.78,109.4) -- (573.78,149.4) -- (506.86,149.4) -- cycle ;
\draw    (501.25,130) -- (506.79,129.92) ;
\draw    (574.22,130) -- (579.76,129.92) ;

\draw (104,90.33) node [anchor=north west][inner sep=0.75pt]  [font=\footnotesize] [align=left] {L1};
\draw (149,90.33) node [anchor=north west][inner sep=0.75pt]  [font=\footnotesize] [align=left] {L1X};
\draw (212,79.33) node [anchor=north west][inner sep=0.75pt]  [font=\footnotesize] [align=left] {BC1};
\draw (368,79.33) node [anchor=north west][inner sep=0.75pt]  [font=\footnotesize] [align=left] {BC2};
\draw (458.67,87.33) node [anchor=north west][inner sep=0.75pt]  [font=\footnotesize] [align=left] {L3};
\draw (294,88) node [anchor=north west][inner sep=0.75pt]  [font=\footnotesize] [align=left] {L2};
\draw (501.33,86.67) node [anchor=north west][inner sep=0.75pt]  [font=\footnotesize] [align=left] {Transport line};
\draw (584,86.67) node [anchor=north west][inner sep=0.75pt]  [font=\footnotesize] [align=left] {Undulators};
\draw (70.67,66.33) node [anchor=north west][inner sep=0.75pt]  [font=\footnotesize] [align=left] {Laser heater};
\draw (45.33,38.33) node [anchor=north west][inner sep=0.75pt]  [font=\footnotesize] [align=left] {CUPID photoinjector};
\draw (181.33,143.73) node [anchor=north west][inner sep=0.75pt]  [font=\scriptsize]
{$ \begin{array}{l}
R_{56} \ =-45.5 \mathrm{mm}\\
T_{566} =\ \ 68.9 \mathrm{mm}
\end{array}$};
\draw (338.67,145.07) node [anchor=north west][inner sep=0.75pt]  [font=\scriptsize]  {$ \begin{array}{l}
R_{56} \ =-28 \mathrm{mm}\\
T_{566} =\ \ 42 \mathrm{mm}
\end{array}$};

\end{tikzpicture}